\documentclass[aps,prl,twocolumn,groupedaddress]{revtex4}

\usepackage{gensymb}
\usepackage{amssymb}
\usepackage{color,graphicx}
\usepackage{subfig}
\usepackage{transparent}
\usepackage{textcomp}
\usepackage{siunitx}
\begin{document}


\title{Resonant charge transfer of hydrogen Rydberg atoms incident at a Cu(100) projected band-gap surface}


\author{J. A.  Gibbard}
\affiliation{Department of Chemistry, University of Oxford, Chemistry Research Laboratory, Oxford OX1 3TA, United Kingdom}
\author{M. Dethlefsen}
\affiliation{Department of Chemistry, University of Oxford, Chemistry Research Laboratory, Oxford OX1 3TA, United Kingdom}
\author {M. Kohlhoff}
\affiliation{Department of Chemistry, University of Oxford, Chemistry Research Laboratory, Oxford OX1 3TA, United Kingdom}
\author{C. J. Rennick}
\affiliation{Department of Chemistry, University of Oxford, Chemistry Research Laboratory, Oxford OX1 3TA, United Kingdom}
\author{E. So}
\affiliation{Department of Chemistry, University of Oxford, Chemistry Research Laboratory, Oxford OX1 3TA, United Kingdom}
\author{M. Ford}
\affiliation{Department of Chemistry, University of Oxford, Chemistry Research Laboratory, Oxford OX1 3TA, United Kingdom}
\author{T. P. Softley}
\affiliation{Department of Chemistry, University of Oxford, Chemistry Research Laboratory, Oxford OX1 3TA, United Kingdom}

\date{\today}

\begin{abstract}

The charge transfer (ionization) of hydrogen Rydberg atoms ($n=25-34$) at a Cu(100) surface 
is investigated. Unlike fully metallic surfaces, where the Rydberg electron energy is degenerate with the conduction band of the metal, the Cu(100) surface has a projected bandgap at these energies, and only discrete image states are available through which charge transfer can take place. Resonant enhancement of charge transfer is observed for Rydberg states whose energy matches one of the image states, and the integrated surface ionization signals (signal versus applied field) show clear periodicity as a function of $n$ as the energies come in and out of resonance with the image states.  The  surface ionization dynamics show a velocity dependence; decreased velocity of the incident H atom leads to a greater mean distance of ionization and a lower field required to extract the ion. The surface-ionization profiles  for `on resonance' $n$ values show a changing shape as  the velocity is changed, reflecting the finite field range over which resonance occurs.

\end{abstract}

\maketitle
The collision of a Rydberg atom in the gas phase with a solid surface typically leads to transfer of the Rydberg electron to the surface at distances less than $5n^2a_0$, where $n$ is the Rydberg electron principal quantum number.  This is especially true for metallic surfaces, where the Rydberg electron energy is degenerate with the conduction band  so that resonant charge transfer (RCT) can occur. Experimental and theoretical studies of this phenomenon have focused on the effects of varying the $n$ quantum number, the parabolic quantum number $k$, the velocity of the incoming particle and the applied fields \cite{eric_prl, wpp_atoms}, and observing how the rate of ionization  varies as a function of distance from the surface \cite{ion_f(d)}. For non-hydrogenic atoms,  adiabatic and non-adiabatic passage through surface-induced energy level crossings leads to behavior that varies with the Rydberg species \cite{dunn_dunh}. Thus, such studies reveal important information about the Rydberg  states and their dynamics near surfaces. \\
\indent
An equally important question for such studies is  what they reveal about the nature of the surface. 
Experimental studies have been primarily conducted with flat-metal surfaces for which the ionization dynamics are almost independent of the material  because of the generic behavior of RCT to the conduction band. 
However, there have also been some experimental and/or theoretical investigations of the effects of adlayers and thin insulating films  \cite{mccown}, interaction with doped semiconductor surfaces \cite{Silicon} and dielectric materials \cite{dielectric}, effects of corrugation and  of patch charges \cite{stray_dunn,dunn_patt}.  Related theoretical calculations were used to investigate  the variation of ionization rate of ground state H\textsuperscript{-} with the thickness of a metal film substrate \cite{H_thin}. 
All these studies point to a degree of sensitivity of the charge transfer process to the surface characteristics. The mean radius of a hydrogenic Rydberg orbit  is of order $n^2a_0$ (e.g., $\sim 20$ nm for $n=20$)  and charge transfer typically occurs at a Rydberg-surface distance of $3-5 n^2a_0$. 
Thus information revealed about  the {\it geometrical} structure of the surface is likely to be limited to nano-scale features. 

In this paper we investigate the RCT of hydrogen Rydberg atoms ($n=25$ to 34) at a Cu(100) surface and  focus on the role that the {\it electronic} structure of the surface plays in the charge transfer, as probed by the resonant nature of the  process. Cu(100) has a band gap at the energy of the Rydberg states and RCT can only occur via `image-charge states'.
An electron outside the surface at a distance $z$ gives rise to an image-charge attractive potential given by (for a perfect conductor),
$V(z)=-\frac{1}{4z}$.
This one-dimensional   Coulomb-like potential can  support an infinite series of bound states forming a Rydberg-type series with energies given by
\begin{equation}
E_{\mathrm{IS}}(n_{\mathrm{img}})=-\frac{R}{16} .\frac{1}{2(n_{\mathrm{img}}+a)^2}
\label{eq:eryd}
\end{equation}
where $n_{\mathrm{img}}$ is the image-state index and $a$ is the quantum defect parameter for a given surface. For Cu(111) $a\approx0.02$ and for Cu (100) $a\approx0.24$ \cite{image_step2}.
Such states are only observable in the bandgap range, as those degenerate with the conduction band are mixed and broadened into the band. 
In the direction parallel to the surface (for both surface and image states)  the wavefunction will be very similar to the bulk metal states 
and energy is not quantized.
In the nearly free-electron model the states form bands with energy
\begin{equation}
\epsilon(k)=E_{\mathrm{IS}}+\frac{\hbar^2}{2m}(k_x^2+k_y^2)
\end{equation}
where $E_{\mathrm{IS}}$ is the energy of the state with zero parallel momentum (Eqn. \ref{eq:eryd}).
There may also be intrinsic surface states in the bandgap; 
whereas for image states the wavefunction is almost entirely outside the metal, for surface states it is located at the surface with some significant penetration inside (similar to the difference between valence and  Rydberg  states of isolated molecules.)
Surface and image charge states have been studied experimentally for a number of materials using  time-resolved two-photon  photoemission (TRPES), inverse photoemission and scanning tunnelling spectroscopy \cite{cu_im_4,STM,invphoem}. 

 The current work extends earlier studies  of charge transfer between ground-state cesium atoms or H$^{-}$ and a Cu(111) surface \cite{Cs_Cu, H-/Cu}, and   studies of the reverse charge transfer to H or Li$^+$ from a Cu (111) surface \cite{hecht}. 
In recent theoretical work we used  a wavepacket propagation method  to determine surface ionization rates versus distance for a moving Rydberg H atom ($n=2-8$) incident at Cu(111) and Cu(100) surfaces \cite{ourtheory}.  We predicted that, for both  surfaces, resonances between the energy of the surface-localized image states and the Rydberg atom result in enhancement of the surface ionization process \cite{ourtheory} such that charge transfer takes place at greater distance from the surface. 
 However the low-$n$ states considered there are not useable experimentally due to their short lifetime with respect to radiative decay. %
Here, we use the long-lived $n=25-34$ states, which fall within the bandgap of the Cu(100) surface; for the Cu(111) surface the projected band gap occurs at an energy below these  Rydberg states.

\begin{figure}
\includegraphics{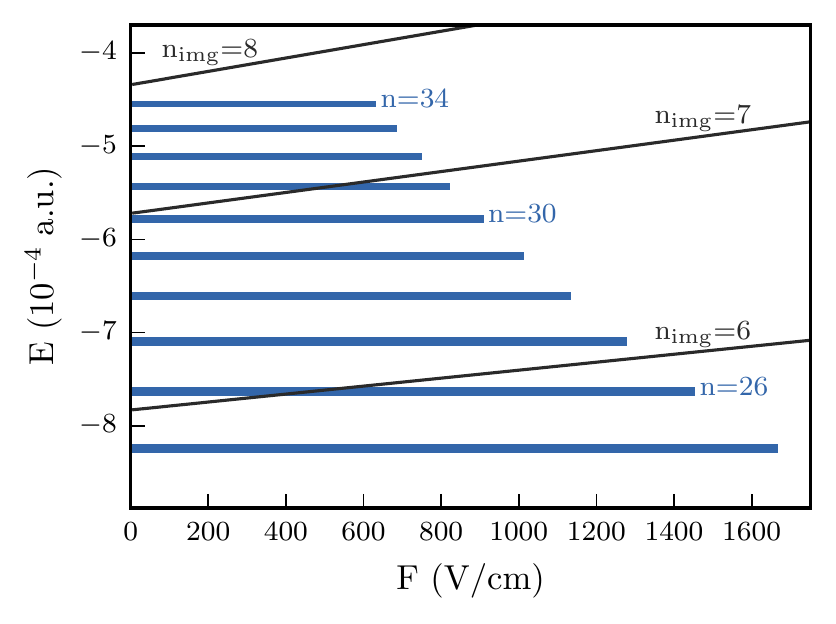}
\caption{The energies of the $n_{\mathrm{img}}=25$ to $n=36$ $k=0$ H atom states, and the image states ($n_{\mathrm{img}}=5$ to $n_{\mathrm{img}}=9$) of the Copper(100)  states. At intersections, resonant enhancement of charge transfer is expected. The widths of the H atom levels lines represent the range of surface perturbation for distances from the surface of $3n^2a_0$ to 6$n^2a_0$}
\label{fig:enlevs}
\end{figure}
Figure \ref{fig:enlevs} shows the predicted  energies of the  $n=25-34$ $k=0$ H-atom  states and the surface-localized image states as a function of applied  field. The parabolic quantum number $k$ runs from $-(n-|m_l|-1)$  and $(n-|m_l|-1)$,  but only
the mid-Stark-manifold $k=0$ Rydberg states are selected for study, as their energies are approximately field independent and provide the greatest region of crossing with the field-dependent image-state energies. 
The energies are calculated by diagonalization of the Hamiltonian using a DVR basis set \cite{lag_leg}.
 The widths of the Rydberg curves shown represent the perturbation of the Rydberg energy due to the surface interaction over the typical range that ionization occurs, $6n^2a_0$ to $3n^2a_0$. Resonance-enhanced charge transfer is expected at applied fields corresponding to the crossing
of the Rydberg and the image states.

\begin{figure}[htb] 
\includegraphics{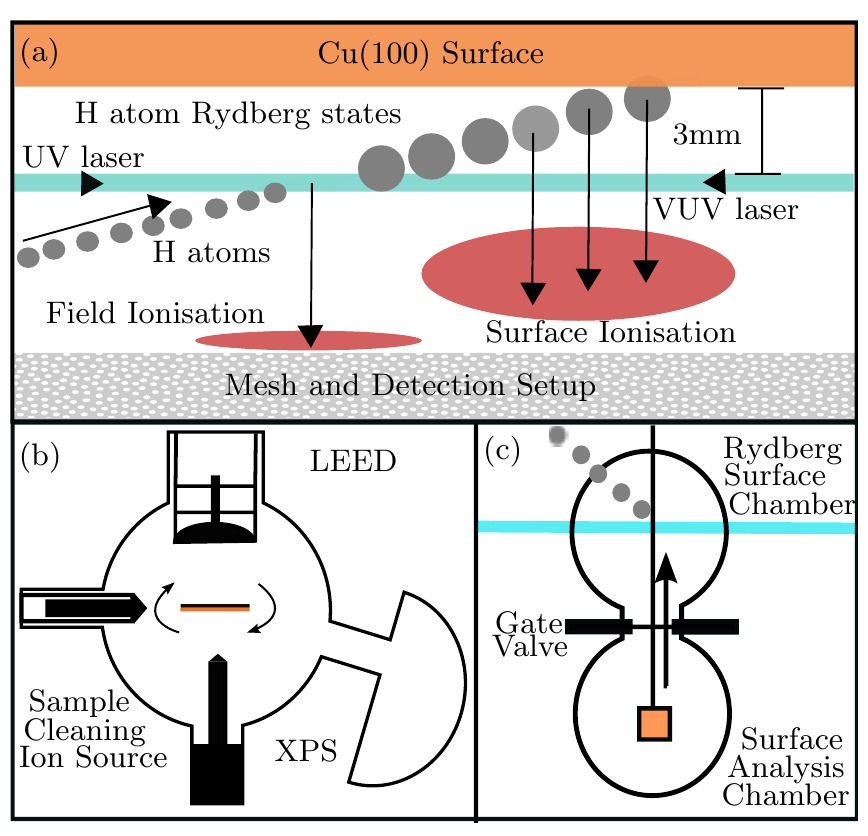}
\caption{(a) Two-color excitation of H atoms produces a beam of Rydberg atoms to probe a Cu(100) surface. Field- and surface-ionization signals are spatially and temporally separated due to the different ionization positions with respect to the surface. (b) Schematic of the surface analysis chamber. (c)  The xyz manipulator allows the surface to be moved between the surface analysis chamber and the Rydberg surface experiment under vacuum. }
\label{fig:expset}
\end{figure}

The experimental setup to measure the H atom--surface interactions was described previously \cite{eric_prl}. In brief, H atoms are formed by photolysis of a supersonic beam of NH$_3$ at $193\si{\nano\meter}$ in a capillary mounted on the pulsed nozzle. Using a pure NH$_3$ beam, the H atoms travel  $50\si{\centi\meter}$  to the laser excitation point where the high-$n$ Rydberg states are populated by two-color ($\lambda_1= 121.57\si{\nano\meter}$, $ \lambda_2=365.75$ to $366.75\si{\nano\meter}$) two-photon excitation via the $2p$ intermediate level. The excitation occurs in a large enough field  to allow selection of a particular Stark state ($k=0$  in this work) of the $n$-manifold. As shown in figure \ref{fig:expset}(a) the Rydberg atoms then travel  $3\si{\milli\meter}$ to interact with the surface which is mounted at a 15\degree \  incidence angle with respect to the atom beam. 
The surface can be moved under vacuum to a surface-analysis chamber, see Fig. \ref{fig:expset}(b), where Low Energy Electron Diffraction (LEED)  and  X-ray Photoelectron Spectroscopy (XPS) are used to determine the elemental composition, crystal plane and presence of impurities on the surface.  
The single crystal Cu(100) surface (Mateck GmbH) was prepared in vacuum using 30 mins. of argon-ion bombardment at 500 eV followed by heating at $700\si{\degree\celsius}$ for 20 mins. Twenty sputtering and annealing cycles produced a clear LEED pattern with the expected four-fold symmetry, and XPS showed only trace oxygen impurities on the surface. This indicated a clean and flat copper(100) surface.  

\begin{figure}
\includegraphics{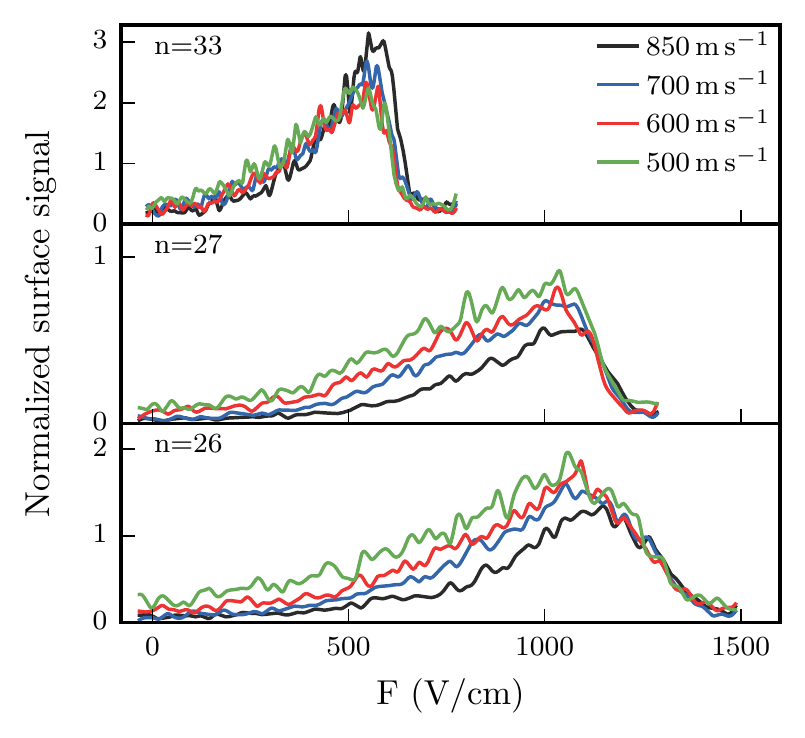}
\caption{Surface ionization profiles for the H atoms with $n= 26$, 27, 33 at various collisional velocities. Black profiles are perpendicular collisional velocity $850\si{\meter \per \second}$, red $700\si{\meter \per \second}$, green $600\si{\meter \per \second}$ and blue $500\si{\meter \per \second}$.} 
\label{fig:sip} 
\end{figure}

Ions, resulting from the Rydberg-to-surface electron transfer, are extracted away from the surface to a detector by applying a field perpendicular to the surface. The field is present at the time of ionization, and is switched from the initial Stark field 1\textmu s after excitation.  
The minimum field required to extract the ions $(\vec{F}_{\rm min})$ depends on the Rydberg-surface separation (D) at which ionization occurs and also on the kinetic energy of the incident H atom along the surface normal $(T_{\perp})$; for a perfect conductor,
\begin{equation}
\label{eq:fmin}
\vec{F}_{\mathrm{min}} (D, T_{\perp}) =\Bigg[ \frac{1}{2D}+\sqrt{\frac{T_{\perp}}{D}} \Bigg]^2
\end{equation}
The mean ionization distance is also mildly dependent on the applied extraction field \cite{wpp_atoms}. Experimentally we observe `surface ionization profiles' as in Fig. \ref{fig:sip}, each recording the intensity of surface ionization signal (ions)  as a function of extraction field. 
The profiles shift to higher field as $n$ decreases, because ionization occurs closer to the surface and a greater field is required to prevent the ion being pulled into the surface by its own image charge and velocity. The gradual rise of the signal as the field increases reflects the range of distances over which ionization takes place, but  also reflects the distribution of surface charges (patch fields) that affect primarily the extraction probability  \cite{ion_range, dunn_patt}.

 At sufficiently large fields the Rydberg atom is field-ionized before reaching the surface, leading to a high-field cut off in the surface-ionization profile. Ions from direct field ionization  are separated in time from the surface ionization signal: the field ionization signal is used to normalize the surface ionization profiles, to account for fluctuations in laser power and molecular beam density. 
 
As discussed below,  measuring surface ionization profiles for different velocities provides an extra handle on resonant charge transfer effects.
Experimentally, the  velocity is varied by changing the delay between the photodissociation laser pulse and the two excitation lasers. Over the $50\si{\centi\meter}$ distance between the photolysis of NH$_3$ and Rydberg excitation of H atoms, the atoms spread out in the longitudinal direction according to their velocity; hence changing the delay picks out a different perpendicular velocity component, variable between  $500\si{\meter \per \second}$ and $850\si{\meter \per \second}$ for an unseeded H atom beam, with a velocity resolution of 1$\%$.

\begin{figure}
\includegraphics{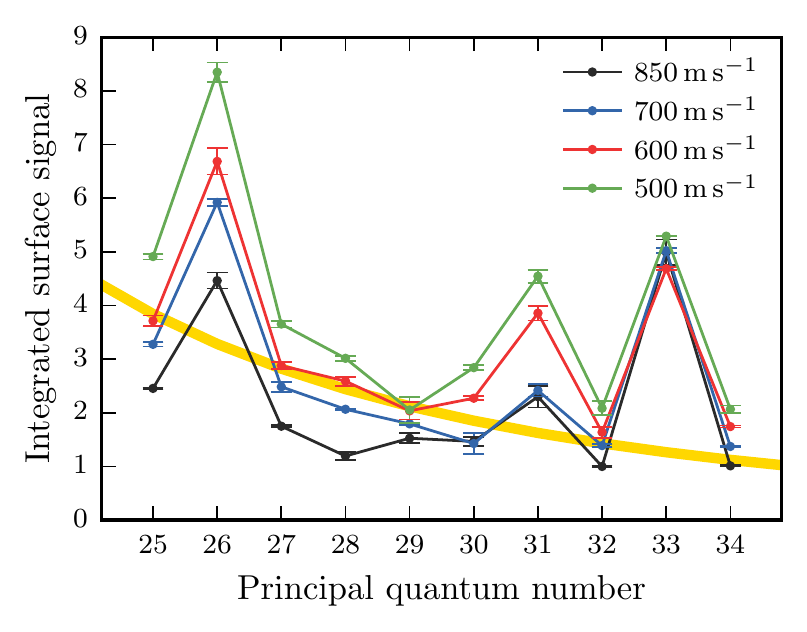}
\caption{The integrated surface signal at a Cu(100) surface as a function of hydrogen principal quantum number. Clear periodicity is seen, indicating resonant enhancement of charge transfer. The yellow line shows the corresponding behavior for a gold surface at a velocity of 660 ms$^{-1}$ } 
\label{fig:intsig} 
\end{figure}

Example surface ionization profiles are shown in Fig. \ref{fig:sip} for three different $n$  values and four different incident velocities, while Fig. \ref{fig:intsig} plots the integral of the surface ionization profile for $n=25-34$.
The profiles in Fig. \ref{fig:sip} appear similar to those measured for  H atoms incident at a gold surface \cite{eric_prl} (for which RCT can occur at all $n$ values
as there is no band gap.)
 Figure \ref{fig:intsig} shows that  
the integrated surface profile for gold (yellow line) decreases monotonically as $n$ increases, because the field range over which ions are extracted decreases with increasing $n$. 
For Cu(100) however there are larger variations in the intensity of the  surface ionization signal as a function of $n$. The maximum surface ionization signal (normalized to field ionization) for $n=26$ is twice that for $n=27$ (Fig. \ref{fig:sip}),  and the low-field part of the profile is also raised in intensity. 
Other such higher intensity profiles are seen at $n=31$ and $n=33$. The increased intensity at lower extraction fields implies a higher propensity for ionization  at a greater distance from the surface.
\indent

In Fig. \ref{fig:intsig} clear resonances can be seen as $n$ is varied, with peaks at $n=26$ and $n=33$, and a smaller peak at $n=31$. 
We attribute this non-monotonic behavior to the predicted resonance effect resulting from energy matching between the Rydberg state and the  image-charge state.
 As shown by theoretical calculations \cite{ourtheory}, in the non-resonant case the charge transfer can only occur if the electron takes up significant momentum parallel to the surface. Conservation of angular momentum inhibits the development of substantial parallel momentum (this is a high angular momentum state with respect to the atom), and there is a marked preference for the electron flux to occur perpendicular to the surface. In addition the saddle point in the electronic potential occurs along the perpendicular direction and hence classically  electron transfer should occur in this direction. 
Fig. \ref{fig:enlevs} predicts that there are 3 image states, $n_{\mathrm{img}}=6$ to 8, crossing the Rydberg energies at $n=26, 31$ and 35, compared to the experimental peaks at $n=26,31$ and 33.
Some degree of discrepancy is expected given several simplifications in  the energy level calculations and overall we consider this to be very good agreement.

Each connected set of points in Fig. \ref{fig:intsig} represents the integrated surface signal for different incident velocities of hydrogen. In general there are two contributing factors to the velocity dependence of the signals; ion-extraction efficiency and ionization distance.  
First,  a larger extraction field is needed to pull the ion away from the surface for higher  initial velocity (Eqn. \ref{eq:fmin}), and the integrated signal decreases with increasing velocity.  
Second, the more slowly moving atoms will have more time to be ionized at greater distances from the surface even though the ionization rate is slow at such distances, and hence the mean ionization distance shifts to a larger value, reducing the minimum extraction field, and again the integrated signal decreases with velocity.  

For the off-resonant $n$ values, the signal amplitude scales up  as the velocity decreases, and the  signal increase is greater at lower-field values - e.g, for $n=27$ the signal at $500\si{\volt\per\centi\meter}$  and  velocity  $500\si{\meter \per \second}$ is $\sim4$ times its value at $850\si{\meter \per \second}$, whereas at $1000\si{\volt\per\centi\meter}$ the difference is only a factor of 1.5.   
For the resonant $n$ values, the signal enhancement occurring as the velocity is lowered tends to lead to a change in shape of the profile.  For example for $n=33$ the $850\si{\meter \per \second}$ signal (black  in Fig. 3) lies above the $850\si{\meter \per \second}$ signal (green) at $550\si{\volt\per\centi\meter}$ but lies below it at $300\si{\volt\per\centi\meter}$.  We believe the shape change  happens because the resonance only occurs in a certain field range, and the velocity effects are likely to be different when the system is in the  resonant field range  compared to the off-resonant field range.  The ionization occurs further from the surface in the resonant range,  and there will be a greater dependence of the observed signals on velocity for more distant ionization \cite{ourtheory}. The second term in Eqn. (3) becomes more important as $D$ increases for a given value of $T$, and hence $F_{\mathrm{min}}$ varies more with $T$ (and hence with collisional velocity.)\\
\indent
In this work we have demonstrated that the predicted resonances between hydrogen atom Rydberg states and the image states within the projected band gap of a copper(100) surface are experimentally observable. 
The resonances occur in particular field ranges corresponding to a range of crossing  between Rydberg and image states. 
Varying the velocity of the incoming beam provides a useful additional diagnostic for the existence of the resonance effects. 
 This work shows that the Rydberg-surface collision experiment can lead to useful information about the electronic structure of the surface, not just the Rydberg atom itself. This type of experiment may be applicable to other systems where there is quantization of the surface states e.g., for thin films or nanostructures, and such surfaces are currently under investigation. The attractiveness of using Rydberg charge transfer arises from the wide range of energies that can be probed by populating different Rydberg quantum states.


\end{document}